\documentstyle[12pt]{article}

\begin{document}
\bibliographystyle{unsrt}

\hoffset -1.5cm
\date{}
\newcommand{\rea}{{\rm I\hspace{-0.5mm}R}}
\newcommand{\nat}{{\rm I\hspace{-0.4mm}l\hspace{-1.1mm}N}}

\newcommand{\comp}{\mbox{\rm\hspace{1.2mm}
\rule[0.05mm]{0.1mm}{2.5mm}\hspace{-1.2mm}C}}
\newcommand{\calk}{\rule[2.2mm]{0.1mm}{0.5mm}\hspace{-0.3mm}{\sf
Z}\hspace{-1.7mm}{\sf Z}\hspace{-0.3mm}\rule{0.1mm}{0.5mm}}


\newcommand{\Top}[1]{\mbox{${\rm top}(#1)$}}
\newcommand{\Sc}[1]{\mbox{${\rm sc}(#1)$}}
\newcommand{\Gen}[1]{\mbox{${\rm Gen}(#1)$}}
\newcommand{\Lc}[2]{\mbox{${\rm lc}_{#1}(#2)$}}

\newcommand{\adres}[1]{  {\small\it  \begin{center} #1 \end{center}   }     }
\newcommand{\dzieki}{ \begin{center}{\bf Acknowledgments} \end{center}  }
\newcommand{\dow}{ \noindent {\bf Proof. }  }
\newcommand{\ckow}[4]{\mbox{$#1_{#2},#1_{#3},\dots ,#1_{#4}$}}
\newcommand{\tildeckow}[4]{\mbox{$\tilde{#1}_{#2},
\tilde{#1}_{#3},\dots ,\tilde{#1}_{#4}$}}
\newcommand{\barckow}[4]{\mbox{$\bar{#1}_{#2},
\bar{#1}_{#3},\dots ,\bar{#1}_{#4}$}}
\newcommand{\ckon}[4]{\mbox{$#1^{#2},#1^{#3},\dots ,#1^{#4}$}}
\newcommand{\tildeckon}[4]{\mbox{$\tilde{#1}^{#2},
\tilde{#1}^{#3},\dots ,\tilde{#1}^{#4}$}}
\newcommand{\barckon}[4]{\mbox{$\bar{#1}^{#2},
\bar{#1}^{#3},\dots ,\bar{#1}^{#4}$}}
\newcommand{\ckk}[7]{\mbox{$#1_{#2}^{#3},#1_{#4}^{#5},
\dots ,#1_{#6}^{#7}$}}
\newcommand{\tildeckk}[7]{\mbox{$\tilde{#1}_{#2}^{#3},
\tilde{#1}_{#4}^{#5},\dots ,\tilde{#1}_{#6}^{#7}$}}
\newcommand{\barckk}[7]{\mbox{$\bar{#1}_{#2}^{#3},\bar{#1}_{#4}^{#5},
\dots ,\bar{#1}_{#6}^{#7}$}}
\newcommand{\ciag}[4]{\mbox{$#1=#2,#3,\dots,#4 $}}
\newcommand{\dsp}[1]{\mbox{$(#1,{\cal #1})$}}
\newcommand{\tildedsp}[1]{\mbox{$(\tilde{#1},\tilde{{\cal #1}})$}}
\newcommand{\bardsp}[1]{\mbox{$(\bar{#1},\bar{{\cal #1}})$}}
\newcommand{\dspa}[2]{\mbox{$(#1,{\cal #2})$}}
\newcommand{\fun}[3]{\mbox{$#1:#2 \- \rightarrow #3$}}

\newcommand{\genb}{\mbox{\ckow{\alpha^\bullet}{0}{1}{4}}}
\newcommand{\geno}{\mbox{\ckow{\alpha^\circ}{0}{1}{4}}}
\newcommand{\poo}{\mbox{$\cal {P}^\circ$}}
\newcommand{\zpo}{\mbox{$ {P}^\circ$}}
\newcommand{\zpto}{\mbox{$ \tilde{P}^\circ$}}
\newcommand{\pto}{\mbox{$ \tilde{\cal P}^\circ$}}
\newcommand{\pbb}{\mbox{$\cal {P}^\bullet$}}
\newcommand{\zpb}{\mbox{$ {P}^\bullet$}}
\newcommand{\zptb}{\mbox{$ \tilde{P}^\bullet$}}
\newcommand{\ptb}{\mbox{$ \tilde{\cal P}^\bullet$}}
\newcommand{\pp}{\dsp{P}}
\newcommand{\tpp}{\tildedsp{P}}
\newcommand{\ppo}{\dsp{P^\circ}}
\newcommand{\tppo}{\tildedsp{P^\circ}}
\newcommand{\ppb}{\dsp{P^\bullet} }
\newcommand{\tppb}{\tildedsp{P^\bullet}}
\newcommand{\psio}{\mbox{${\Psi}^{\circ}_{\epsilon,l,\beta}$}}
\newcommand{\psib}{\mbox{${\Psi}^{\bullet}_{\epsilon,l,\beta}$}}
\newcommand{\psitb}{\mbox{$\tilde{\Psi}^{\bullet}_{\epsilon,l,\beta}$}}
\newcommand{\psito}{\mbox{$\tilde{\Psi}^{\circ}_{\epsilon,l,\beta}$}}
\newcommand{\betab}{\mbox{${\beta}^{\bullet}$}}
\newcommand{\phb}{\mbox{$ \hat{\cal P}^{\bullet}$}}
\newcommand{\hatppb}{\mbox{$\dspa{\zpb}{\phb}$}}
\newcommand{\cbdo}{\mbox{$\Box$}}
\newcommand{\czero}{\mbox{${\cal C}_0$}}
\newcommand{\tao}{\mbox{$\tilde{A}^{\circ}$}}
\newcommand{\ao}{\mbox{$A^{\circ}$}}
\newcommand{\tab}{\mbox{$\tilde{A}^{\bullet}$}}
\newcommand{\ab}{\mbox{$A^{\bullet}$}}
\newcommand{\boldtao}{\mbox{$\tilde{\bf A}^{\circ}$}}
\newcommand{\boldao}{\mbox{${\bf A}^{\circ}$}}
\newcommand{\boldtab}{\mbox{$\tilde{\bf A}^{\bullet}$}}
\newcommand{\boldab}{\mbox{${\bf A}^{\bullet}$}}
\newcommand{\piro}{\mbox{${\pi}_{\rho _{_{H}}}$}}
\newcommand{\tpiros}{\mbox{$\tilde{\pi}_{\rho _{_{H}}}^{\ast}$}}
\newcommand{\piros}{\mbox{${\pi}_{\rho _{_{H}}}^{\ast}$}}
\newcommand{\piroh}{\mbox{${\pi}_{\rho _{_{H}}}^{\#}$}}


\newtheorem{tw}{Theorem}[section]
\newtheorem{lem}{Lemma}[section]
\newtheorem{stw}{Proposition}[section]
\newtheorem{defin}{Definition}[section]
\newtheorem{wn}{Corollary}[section]
\newtheorem{rysunek}{Figure}[section]




\title{\bf Differential Structure of Space-Time and Physical Fields}
\author{Jacek Gruszczak \thanks{email: sfgruszc@cyf-kr.edu.pl} }
\maketitle
\adres{Institute of Physics, Pedagogical University, ul. Majora 41/7,
       31-422 Cracow,Poland}

\vspace{1cm}

\vspace{1cm}


\begin{abstract}
A physical interpretation of axioms of the differential structure of
space-time is presented. Consequences of such interpretation for
cosmic string's space-time with a scalar field are studied. It is shown
that the
assumption of smoothness of the scalar field leads either to modification of
cosmic string's space-time global properties or to quantization of
the deficit of angle: $\Delta =2\pi (1-1/n)$, $n=1,2,\dots$.

\end{abstract}
\vspace{2cm}
Lecture delivered at "Workshop on Modern Methods in Classical and
Quantum Gravity" , July 26-28th, 1995, Sintra, Portugal.


           \section{Introduction}
                              \label{intro}

             There  are  various  ideas  of  generalization of the manifold
          notion.   For  example  Aronszajn  and  Marshall
          \cite{1af,27af}
          have developed theory of so-called subcartesian spaces.
             Mostov  \cite{28af}  and Spallek \cite{36af,59af}
          have  defined  spaces  called at present Mostov spaces.
           Roman Sikorski
          defined  differential  spaces  (d-spaces for short)
          \cite{3bo}.

             There is a
          strong requirement for notions more general than manifolds
          in  theoretical  physics both classical and quantum .
             Here are few examples.
  \begin{itemize}
   \raggedright
   \item[   a)] Theory  of  singularities in General Relativity; space-time
            with  its  singular boundary is not a manifold
            \cite{33af,1bo},
   \item[   b)] The space of solutions of Einstein equation (superspace) is
            not  a  manifold \cite{12bo},
   \item[   c)] A  class of non-differential processes described in book by
            Arnold \cite{13bo}.
   \end{itemize}
             The  generalizations  of  the  manifold concept have not found
          applications  in  physics yet. Among reasons of such situation I
          can   mention    excessive  abstractness  of  some
          theories and  small  amount  of  theorems useful in practice.

             Differential   spaces   in   the   sense   of   Sikorski   are
          generalization  of the manifold concept by omitting the condition
          of  existence of local diffeomorphisms to ${\rea}^n$. The
          generalization
          admit  to  considerations  larger class of object than manifolds.
          Simultaneously,  mathematical tools (definitions and theorems) of
          the  d-spaces  theory  is  similar  to  that  one usually used in
          ordinary   differential   geometry.
          From the point of view of
          physics,  where  differential geometry plays a very important role,
          differential   geometry   on  d-spaces  is  a  natural  tool  for
          generalization  of  a  physical  theory. Other important argument
          which distinguishes the d-spaces theory is a great number of useful
          theorems.

             Among  various  structures  existing  in  a  space-time  (e.g.
          topological,  chronological,  metrical, spinorial etc. )  only
          differential   structure   is  not  well  elaborated.  Since  the
          structure  may play a fundamental role in quantization of gravity
          \cite{14bo}
          every result
          concerning d-structure of space-time models seems to be valuable.

         \pagebreak

         \section{Differential spaces in the sense of Sikorski}
                               \label{ds}


A differential space in the sense of Sikorski is defined as follows.

Let $(M,{\rm top}(M))$ be a topological space and \czero \
a set of real functions on $M$.
A function \fun{f}{M}{\rea} is local \czero -function on $M$ if for every
$p_0\in M$ there is a neighbourhood $U\in{\rm top}(M)$  and
$\varphi\in\czero$ such that $f|_U=\varphi|_U$.
The set of all local  \czero \ functions on $M$ is denoted by
$(\czero)_{_M}$.
\begin{defin}
A set of real functions ${\cal C}$ on $M$ is said to be closed with
respect to localization if ${\cal C}=({\cal C})_{_M}$.
\end{defin}
\begin{defin}
A pair \dspa{M}{C} \ is said to be a differential space in the sense of
Sikorski if:
\begin{itemize}
\label{defds}
\raggedright
\item[1)] ${\cal C}$ is closed with respect to localization, i.e.
${\cal C}=({\cal C})_{_M}$,
\item[2)] ${\cal C}$ is closed with respect to superposition with smooth
functions on ${\rea}^n$, $n\in\nat$, i.e. for any
$\ckow{\alpha}{1}{2}{n}\in {\cal C}$ and any
$\omega\in C^{\infty}(\rea^n)$, $\omega\circ
(\ckow{\alpha}{1}{2}{n})\in {\cal C}$.
\item[3)] $M$ is equipped with the topology ${\tau}_{_{\cal C}} $  which
is the weakest topology on $M$ in which functions of ${\cal C}$ are
continuous.
\end{itemize}
\end{defin}
A picture of our world is formed by measures providing us with
information coded in form of real functions. One can say that reality is
given to us by a family of real functions existing on $M$.
If the family is sufficiently "complete"  one can identify it with the
 d-structure ${\cal C}$. Then the first axiom of definition \ref{defds}
 guarantees consistency of a local physics with the global one. The second
axiom  means that information obtained by superposition with smooth functions
on
 ${\rea}^n$ do not lead to a new information not contained in ${\cal C}$
 (see \cite{13af}).

Let us  test how the above
described idea works in practice on example of cosmic string's space-time
with a scalar field.

\pagebreak


            \section{D-space of cosmic string's space-time}
                             \label{cosmic}


The space-time of cosmic string $(M,g)$ is a pseudoriemannian manifold
with conical type quasiregular singularity and equipped with the metric
\begin{equation}
g=-dt^2+k^{-2}d\rho^2+{\rho}^2 d\phi^2 +dz^2 ,       \label{metric}
\end{equation}
where $k=(1-\Delta/2\pi)$ , $t, z \in \rea$, $\rho\in (0, \infty)$,
$\phi\in \langle 0, 2\pi ) $ and $\Delta $ stands for
 the deficit of angle. This manifold
is isometric to $(C^{\circ}\times \rea^2, \iota^*\eta^{(5)})$, where
$C^\circ$ is two-dimensional cone without vertex,
$\iota : C^{\circ}\times \rea^2$ $ \rightarrow\rea^5$ is an  embedding and
${\eta}^{(5)}$  is the five-dimensional Minkowski metric.

The singular boundary in this approach is represented by the set
${\rm S}\times \rea^2$, where ${\rm S}$ denotes vertex of the cone
$C^{\bullet}$ and the dot under $C$  denotes that now the vertex is
taken into account.
The presented dipheomorphic picture of a space-time of cosmic string is
interesting for two reasons. First, it shows
the sense of the conical singularity in a demonstrative
manner \cite{16af,15af}. Second, it enables
us to construct immediately the differential structure
in the sense of Sikorski for
cosmic string's space-time both with and without singularity.

In this approach, d-space of a cosmic string constitutes d-subspace
$(C^{\circ}\times \rea^2,$ $ ({\cal E}_5)_{C^{\circ}\times \rea^2})$
of
the d-space $(\rea^5, {\cal E}_5 )$,
where
${\cal E}_5 =C^{\infty} (\rea^5)$
and the symbol  $(\cdot)_{C^{\circ}\times \rea^2}$ denotes an operation
of taking closure with respect to localization (definition
\cite{3bo,12af,62af}).

The cosmic string's space-time with singularity is not a manifold but it is
still a d-space in the sense of Sikorski. The set
$C^{\bullet}\times \rea^2 \subset \rea ^5$ can be treated as  a d-subspace
of
the d-space $(\rea^5, {\cal E}_5 )$. Then the d-space
$(C^{\bullet}\times \rea^2, ({\cal E}_5)_{C^{\bullet}\times \rea^2})$
 represents background d-space of cosmic string's space-time with
 singularity.

The above mentioned d-spaces of cosmic string are not convenient from
technical and methodological point of view. Therefore, in the following,
two auxiliary d-spaces \tppo \ and \tppb \ will be investigated rather
than
$(C^{\circ}\times \rea^2, ({\cal E}_5)_{C^{\circ}\times \rea^2})$
and
$(C^{\bullet}\times \rea^2, ({\cal E}_5)_{C^{\bullet}\times \rea^2})$,
respectively.
The auxiliary d-spaces are defined as follows.

\begin{defin}
Let the set
$\tilde{P}^\circ:=\rea\times
(0,\infty)\times \langle 0,2\pi\rangle \times \rea$ be a "parameter
space".
One can define the following d-structure
$\pto :={\rm Gen}(\ckow{\tilde{{\alpha}}^{\circ}}{0}{1}{4})$,
where functions
\fun{{\tilde\alpha}^{\circ}_{i}}{\tilde{P}^\circ}\rea, $i=0,1,\dots ,4$
are given by means of the following formulas
\begin{quote}
\raggedright
   ${\tilde\alpha}^{\circ}_{0}(\tilde p):=t$ \linebreak
   ${\tilde\alpha}^{\circ}_{1}(\tilde p):=\rho\cos\phi$  \linebreak
   ${\tilde\alpha}^{\circ}_{2}(\tilde p):=\rho\sin\phi$\linebreak
   ${\tilde\alpha}^{\circ}_{3}(\tilde p):=z$\linebreak
   ${\tilde\alpha}^{\circ}_{4}(\tilde p):=\rho$,
   $\tilde{p}\in \tilde{P}^\circ $.
\end{quote}
\end{defin}
The pair \tppo \ is a d-space useful for description of interior of
cosmic string's space time.

\begin{defin}
Let
$\tilde{P}^\bullet:=\rea\times
\langle 0,\infty)\times \langle 0,2\pi\rangle \times \rea$,
be the "prolonged parameter space".
One can define the following d-structure
$\ptb :={\rm Gen}(\ckow{{\tilde{\alpha}}^{\bullet}}{0}{1}{4})$.
The functions
\fun{\tilde{\alpha}^{\bullet}_{i}}{\tilde{P}^\bullet}{\rea} are
defined as follows
$$\tilde{\alpha}^{\bullet}_{i}(\tilde{p}_\bullet):=
\lim_{\tilde{p}\rightarrow\tilde{p}_\bullet}
{\tilde\alpha}^{\circ}_{i}(\tilde{p}), $$
where $\tilde{p}\in\tilde{P}^\circ,\tilde{p}_\bullet\in
\tilde{P}^\bullet$ and $i= 0,1,\dots ,4$.
\end{defin}
The \tppb \ is used for description of a cosmic string's space time with
singularity.

Both \tppo \ and \tppb \ are not Hausdorff topological spaces. Therefore
they are not dipheomorphic to
$(C^{\circ}\times \rea^2, ({\cal E}_5)_{C^{\circ}\times \rea^2})$
and
$(C^{\bullet}\times \rea^2, ({\cal E}_5)_{C^{\bullet}\times \rea^2})$,
respectively.
However, after some identifications of points of \zpto \ and of \zptb \
one can obtain d-spaces dipheomorphic to above mentioned background
d-spaces of cosmic string's space-time (see \cite{16af,15af,62af} for
details). In addition, every statement concerning smoothness and
differential dimension true for \tppo \ and \tppb \ holds also for
$(C^{\circ}\times \rea^2, ({\cal E}_5)_{C^{\circ}\times \rea^2})$
and
$(C^{\bullet}\times \rea^2, ({\cal E}_5)_{C^{\bullet}\times \rea^2})$
(see \cite{62af} for details). Thus, without loss of correctness one can
confine further considerations to \tppo \ and \tppb .


 \section{Physical fields and differential structure for cosmic string}
                            \label{scfield}

Let us consider normal modes of a Klein-Gordon field on the cosmic
string's space-time background. The modes have the following form:

$$\fun{\tilde{\Psi}^{\circ}_{\epsilon,l,\beta}}{\tilde{P}^{\circ}}{\comp},$$
\begin{equation}
\label{psito}
\tilde{\Psi}^{\circ}_{\epsilon,l,\beta}(t,\rho,\phi,z)=
N_{\epsilon,l,\beta}e^{-i\epsilon t}e^{i\beta
z}e^{il\phi}\rho^{|l|/k}F(l,k; \rho),
\end{equation}
where
$\epsilon,\beta\in \rea$,
$l\in\calk$,
$N_{\epsilon,l,\beta}$
is the normalization constant and $k\in (0, 1) $ is defined in
formula (\ref{metric}). The $F$ is an analytical function. Its detailed
form is without meaning for further studies.
\begin{defin}
Let \dspa{M}{C} \ be a d-space. A complex function \fun{f}{M}{\comp} \
is said to be  smooth complex function if \ ${\rm Re}f$  and ${\rm Im}f$
are smooth ones.
\end{defin}

It is easy to check that
\begin{stw}
\label{smootho}
 For every
$\epsilon, \beta\in\rea$,  $l\in\calk$ and $k\in (0, 1)$
the normal modes
\psito \
are smooth functions on \tppo.
\end{stw}

\begin{stw}
\label{smoothb}
For every $\epsilon ,\beta \in \rea$ and $l\in \calk$ the
 modes naturally prolonged to singularity
\begin{equation}
\label{psitb}
\psitb (p):=\lim_{q\rightarrow p}\psito (q), q\in\zpto, p\in\zptb,
\end{equation}
are
\begin{itemize}
\raggedright
\item[{\bf a)}]
 smooth functions on \tppb \ for $k=1/n$, $n=1,2,\dots$,

\item[{\bf b)}]  not smooth functions on \tppb \ for other $k\in (0,1)$.
\end{itemize}
\end{stw}

Following proposition \ref{smootho}
 one can say that
 the d-space of a cosmic string's gravitational field is
from the very beginning "prepared"  for insertion of a Klein-Gordon field
on its background. The fact of insertion of the Klein-Gordon field is from
the point of view of cosmic string's space time interior the only an
indication which functions among already existing in \pto \ are the normal
modes.

The situation is much more interesting when one takes into account
the conical singularity (proposition \ref{smoothb}). Then, in general,
the insertion
of the scalar field enforces  a supplementation of \ptb .
The situation clarifies the following proposition.

\begin{stw}
Let $\hat{\cal P}^{\bullet}$ be the smallest d-structure containing
\psitb \ as smooth functions;
$\hat{\cal P}^{\bullet}:= {\rm
Gen}(\ckow{\tilde{\alpha}^\bullet}{0}{1}{4},\psitb )$. Then the d-space
$(\zptb ,\phb )$ has the following properties
\begin{itemize}
\item[{\bf a)}]
$(\zptb ,\phb ) = \tppb$ for $k=1/n$, $n=1,2,\dots$,
\item[{\bf b)}]
$(\zptb ,\phb )$ \ is not dipheomorphic to \tppb \ for $k\ne 1/n$,
$n=1,2,\dots$,
\item[{\bf c)}]
{\rm dim} $T_{p}(\zptb,\phb) = 6$ and {\rm dim} $T_{p}\tppb = 5$
for $p\in {\bf S}$, where {\bf S} denotes the set of singular points.
\end{itemize}
\end{stw}
Details of proofs one can find in \cite{62af} .

For $\Delta\ne 2\pi (1-1/n)$   the insertion of a Klein-Gordon field
changes global properties of cosmic string's background d-space. The new
background d-space  $(\zptb,\phb)$ is not dipheomorphic to \tppb . For
example the dimension of its tangent spaces at singular points is 6
while for \tppb \ is 5. That means that $(\zptb,\phb)$ do not represent
 background d-space of a cosmic string.

In the case of $\Delta = 2\pi (1-1/n)$ the supplemented d-structure \phb
\ and \ptb \ are the same so the background d-space of a cosmic string's
space-time does not change.

The similar result holds for an electromagnetic field and gravitational
radiation on cosmic string's space-time background \cite{62af}.
Thus, one can formulate the following theorem.

\begin{tw}
\label{nostringA}
The conical space-time of cosmic string equipped with the metric
{\rm  (}\ref{metric}{\rm )} can be a background for
a smooth Klein-Gordon field,  electromagnetic field  and
 a  smooth field of gravitational radiation  the only
in the case of the following discrete spectrum of the deficit of angle:
$$\Delta=2\pi (1-1/n), $$
where $n=1,2,\dots  $\ .
\end{tw}


         \section{Comparison with results obtained by means of
                           field theory methods}
                           \label{gradiation}

The issues of previous section were obtained by means of strictly
geometrical methods within the theory of differential spaces in the
sense of Sikorski. There appear the question: whether an echo of these
anticipations one can find among results obtained by the field theory
methods?

In the present section I will discuss some results of papers by Hacyan
and Sarmiento \cite{44af} and also by Aliev and Gal'tsov
\cite{43af,56af}.

{\bf 1) }
Sarmiento and Hacyan have studied
the energy density spectrum of vacuum surrounding cosmic
string for massless fields with spin 0, 1/2, and 1. Briefly speaking
they have shown that the energy density of vacuum
is given by
the following formula
\begin{equation}
\label{vacuum}
\frac{de}{d\omega}=\frac{\hbar {\omega}^{3}}{2{\pi}^{2}c^{3}}\left(1-
\frac{1}{2\pi\omega k\rho}\sin (\frac{\pi}{k}){\rm
I}_{_{\frac{1}{k}}}(2\omega\rho)\right),
\end{equation}
where $k$ has meaning as in  (\ref{metric}).
When
$k=1/n$, $n=1,2,\ldots$ and so for deficits of angle $\Delta=2\pi (1-1/n)$
the energy density is like for the Minkowski space-time:
$$\frac{de}{d\omega}=\frac{\hbar {\omega}^{3}}{2{\pi}^{2}c^{3}}    . $$
One can say that in this case the energy density is not distorted.
The result is independent of spin.

{\bf 2)  }
The conical nature of space-time cosmic string geometry  is a cause of
a class of effects absent in the Minkowski space-time (for example: the
lensing effect \cite{51af}). One of the most interesting effect
is a gravitational
radiation emitted by a freely moving particle (conical bremsstrahlung).
 The total energy emitted by the particle in the form
of gravitational radiation is given
by the formula
\begin{equation}
\label{radiation}
E=\frac{1}{k}{\sin}^{2}(\frac{\pi}{k})f(v, k, d),
\end{equation}
where $f$ \  is a function of parameters of motion. Its explicit form one can
find in the article by Aliev \cite{43af}. It is easy to check that
the conical bremsstrahlung effect vanishes when
$k=1/n$, $n=1,2,\ldots $ like in the above section.


                        \section{Final remarks}


Results concerning cosmic string's space-time obtained in previous
sections could be interpreted as follows. The assumption of smoothness of
physical fields introduces a new kind of coupling. Usually the coupling
of a gravitational field and physical fields holds throughout the
energy-momentum tensor in Einstein equations. The new coupling has a
global nature and carries out through differential structure of cosmic
string leading to changes in  global properties of this
space-time.
When one enforce  vanishing of such "differential coupling" then
the gravitational field of a cosmic string becomes quantized and
simultaneously every effect mentioned in the section \ref{gradiation}
vanishes.

\end{document}